\documentclass[aps,reprint,superscriptaddress,pra,twocolumn,showpacs]{revtex4-1}

\usepackage{epsfig}
\usepackage{amsmath}
\usepackage{amssymb}
\usepackage{color}
\usepackage{bm}

\newcommand{\ket}[1]{\left|#1\right\rangle}
\newcommand{\bra}[1]{\left\langle#1\right|}
\usepackage{verbatim}
\definecolor{purple}{RGB}{160,32,240}
\definecolor{seegroen}{rgb}{0,0.5,0.5}

\begin{document}

\title{Unitary Equivalence of Quantum Walks}

\author{Sandeep K \surname{Goyal}}
\email{sandeep.goyal@ucalgary.ca}
\affiliation{School of Chemistry and Physics, University of KwaZulu-Natal, Private Bag X54001, 4000 Durban, South Africa}  

\author{Thomas \surname{Konrad}}
\affiliation{School of Chemistry and Physics, University of KwaZulu-Natal, Private Bag X54001, 4000 Durban, South Africa}  
\affiliation{National Institute for Theoretical Physics (NITheP), KwaZulu-Natal, South Africa}

\author{Lajos \surname{Di\'osi}}
\affiliation{ Wigner Research Centre for Physics, Institute for Particle and Nuclear Physics, H-1525 Budapest 114, P.O.B. 49, Hungary} 
\begin{abstract}
A {\em simple} coined quantum walk in one dimension can be characterized by a $SU(2)$ operator with three parameters which represents the coin toss. However, different such coin toss operators lead to equivalent dynamics of the quantum walker. In this manuscript we present the unitary equivalence classes of quantum walks and show that all the nonequivalent quantum walks can be distinguished by a single parameter. Moreover,  we argue that the electric quantum walks are equivalent to quantum walks with time dependent coin toss operator.

\end{abstract}
\maketitle

\section{Introduction}

Propagation of a walker by a succession of random steps based on the outcomes of a coin toss is called classical random walk. In the quantum analogue of this propagation, which is known as discrete time quantum walk, the probability distribution for the coin toss outcomes  is replaced by probability amplitudes of a two-level system -- the quantum coin. Both quantum walk (QW) and classical random walk involve a conditional shift (propagation) of the walker depending on the state of the coin,  but in a QW the superposition of states allows interference of different paths, leading to strikingly different output distributions. For example, the spread of the probability distribution for the  quantum walker increases proportional to the number $n$ of steps as opposed to $\sqrt{n}$ for its classical counterpart \cite{Aharonov1993,Nayak2000,Ambainis2001,Kempe2003,Chandrashekar2008}.

This speedup of QWs promises advantages when applied in quantum computation for certain classes of quantum algorithms \cite{Ambainis2003}, for example, quantum search algorithms \cite{Shenvi2003,Childs2004,Lovett2010}. Moreover, it was shown recently that universal quantum computation can be implemented by means of QWs \cite{Childs2004,Lovett2010}. Quantum walks have also been used to analyze energy transport in biological systems \cite{Mohseni2008}.

In general QW dynamics depend on a number of parameters which define the coin toss operator, an external force acting on the walker,  or varying coin toss operations.  In order to understand and characterize these dynamics one needs  to study in principle the entire parameter space. Which can be a very tedious task. However, sometimes different points in the parameter space result in dynamics which are the same up to a unitary transformation, i.e., a change of basis of the Hilbert space of the system. Such unitarily equivalent dynamics possess qualitatively the same properties, such as the energy levels of the underlying Hamiltonian, the bias of the coin operation, or  the evolution of entanglement between the coin  and the walker etc.. Therefore, it suffices to study only a single example of  dynamics within each equivalence class. 

 For example, for QWs in the presence of a single or several boundaries which can absorb the walker, it was already proven \cite{Bach2004} that the absorption probability can be obtained from considering coin-toss matrices with real entries only. This result was quoted, for example in \cite{Knight2004},  to justify that it is sufficient to consider real-valued coin-toss matrices in studies of ordinary one-dimensional QWs. We here give a conclusive alternative argument showing that indeed the study of  real-valued coin-toss matrices suffices  under certain conditions.    

In this article we present the equivalence classes  for a broad family of one-dimensional QWs defined by a SU(2) coin operator together with a conditional shift of the walker. We  first study the simplest QW dynamics - with time-independent coin toss operation and without external force or noise.  This family of QWs is characterized by  three real parameters which also define the coin-toss operator acting on the coin space. We show that  QW can be reduced to a single-parameter family of QWs and that this parameter determines the group and the phase velocity of the walker.  In \cite{Meyer1996B} Meyer studied  the unitary equivalent classes for one particle quantum lattice gases and obtained the single-parameter family isomorphic to with ours. Afterwards we employ the theory developed to show that QW with step-dependent coin toss is equivalent to the so-called electric QW  \cite{Meyer1996,Wojcik2004,Romanelli2005,Genske2013,Cedzich2013}. In a different way, this has also been shown in \cite{Banuls2006}.

This article is structured as follows: we start by introducing various types of QWs and their parametrizations in Sec.\,\ref{qw-intro}. In Sec.\,\ref{unitary-equiv} we define the unitary equivalence of QW dynamics which results for simple QW in a reduction to a single real degree of freedom (Sec.\,\ref{equiv-class}). In Sec.\,\ref{time-electric-qw} we show that time dependent coined QWs  are equivalent to electric QWs. We conclude with Sec.\,\ref{conc}.

\section{Quantum walks: an introduction}\label{qw-intro}
Let us represent the canonical basis vectors of the coin space by $\{\ket{\uparrow},\ket{\downarrow}\}$ and the  basis of the position space as $\{\ket{j}\}_{j=-\infty}^{\infty}$. The quantum process that resembles the coin toss operation is achieved by realizing a weighted superpositions of heads ($\ket{\uparrow}$) and tails ($\ket{\downarrow}$) as:
\begin{align}
\ket{\uparrow} &\to \alpha \ket{\uparrow} - \beta^*\ket{\downarrow},\\
\ket{\downarrow} &\to  \beta\ket{\uparrow} + \alpha^*\ket{\downarrow},
\end{align}
where $\alpha$ and $\beta$ are complex numbers such that $|\alpha|^2+|\beta|^2=1$. This mapping corresponds to a unitary transformation $U$ which acts on the  two-dimensional space of coin states. 

The \emph{coin toss operator} $U$ can be any unitary operator from the group $SU(2)$:
\begin{align}\label{rotvec}
U = \exp\left(i\frac{\phi}{2}\, \vec{r}\cdot\vec{\sigma}\right),
\end{align}
where $\phi,\vec{r}$ are, respectively, the rotational angle and the unit vector of rotation axis;
 $\vec{\sigma} = (\sigma_x\,\sigma_y\,\sigma_z)$ is the vector of the three Pauli matrices.
 $U$ can be expressed  in Euler decomposition as well:
\begin{align}\label{euler}
U &= \exp\left(i\frac{\eta}{2}\sigma_z\right)\exp\left(i\frac{\theta}{2}\sigma_y\right)\exp\left(i\frac{\xi}{2}\sigma_z\right).
\end{align}
Hence, the coin toss is characterized by three real parameters,  the Euler angles,  $\eta,\,\theta,\,\xi$ or,
alternatively, by the three components of the rotation vector $\phi\,\vec{r}$ \eqref{rotvec}.

As with classical random walk, QW requires a conditional shift of the walker to the left (right) for tails (heads), 
represented by the \emph{conditional shift operator}
\begin{align}
S &= R \otimes \ket{\uparrow}\bra{\uparrow} + R^\dagger\otimes \ket{\downarrow}\bra{\downarrow},
\label{shift}
\end{align}
where $R=\sum_j \ket{j+1}\bra{j}$ and $R^\dagger$ are the right and left translation operator, respectively. The unitary operator $S$ shifts the walker by one position to the left and in superposition to the right depending on the amplitudes of the coin state with respect to the canonical basis $\{ \ket{\uparrow}, \ket{\downarrow}\}$. 

\begin{figure}
\includegraphics[width=7.5cm]{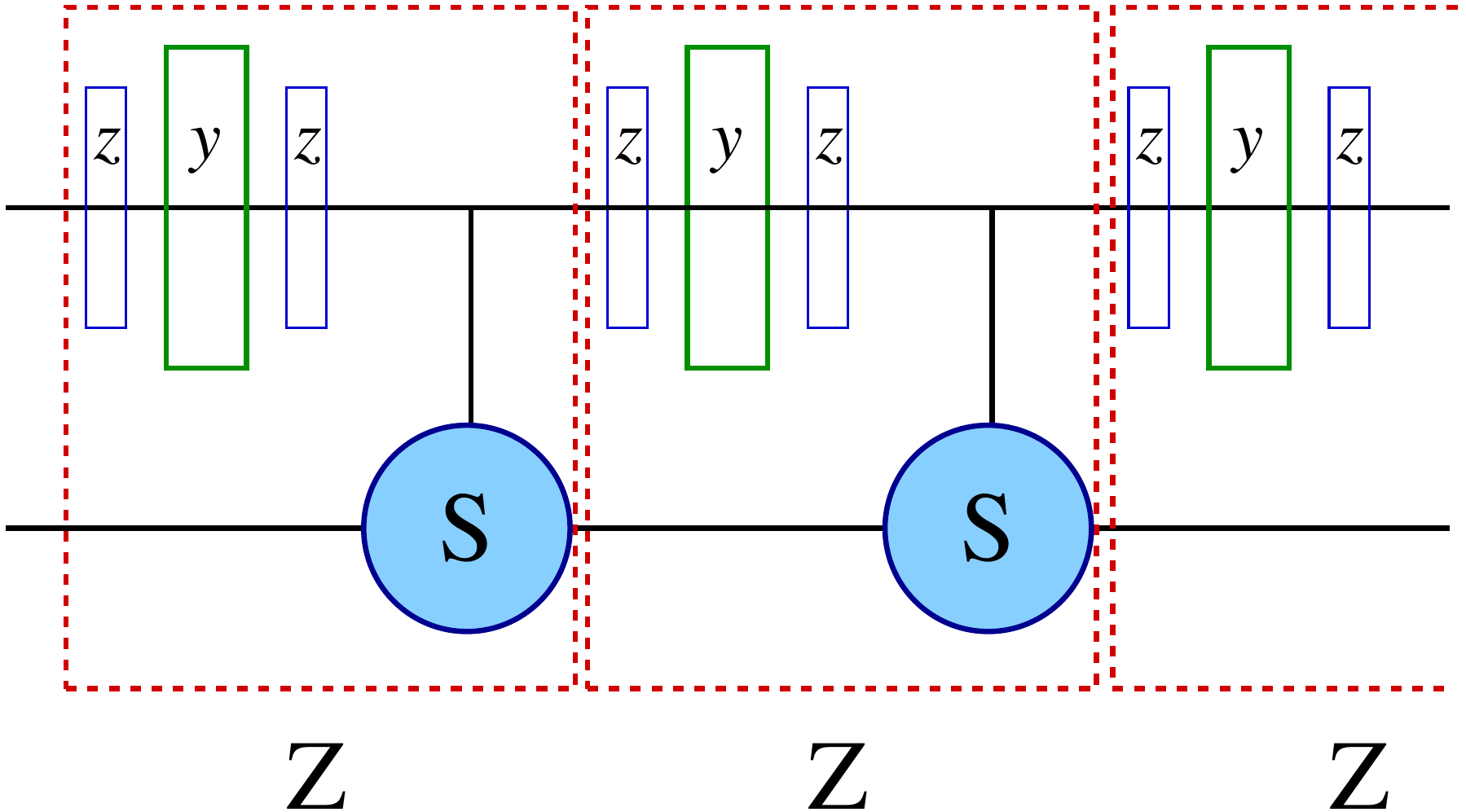}
\caption{(Color online) The circuit diagram of simple QWs. The two horizontal lines represent the coin qubit and the walker qudit ( a superposition of finitely many positions).  The  combination of letters $z,\,y$ and $z$ stand for the three rotations of the coin toss operator $U$ in the Euler representation,  and $S$ is the conditional shift operator. Thus, the dashed block (red)  illustrates the propagation operator $Z$. Repeated action of $Z$ on the initial state of the walker and coin results in the simple QW dynamics.}\label{qw}
\end{figure}

The unitary \emph{propagation operator}  $Z$ of a simple  QW \footnote{By simple we mean a translationally invariant and time-independent quantum.}  is fully determined by the unitary rotation $U$ times the conditional shift $S$  (cp. Fig.\,\eqref{qw}):
\begin{align}\label{Z}
Z &= (\mathbb{I}\otimes U)S,\nonumber\\
&= R \otimes U\ket{\uparrow}\bra{\uparrow} + R^\dagger\otimes U\ket{\downarrow}\bra{\downarrow}.
\end{align}
The repeated action of the operator $Z$ on an initial composite state $\ket{\Psi(0)}$ of the walker and the coin gives rise to QW evolution:
\begin{align}
\ket{\Psi(n)} &= Z^n \ket{\Psi(0)}.
\end{align}
Note that the initial state is usually chosen to be uncorrelated:
\begin{align}
\ket{\Psi(0)}  &= \ket{\psi(0);\mathrm{walker})}\otimes\ket{\chi(0);\mathrm{coin}}.
\end{align}

 We note that the operator $Z$ satisfies the relation 
\begin{align}\label{Z-translation-invariance}
Z =  \left( R\otimes \mathbb{I}\right) Z  \left(R^\dagger\otimes \mathbb{I}\right).
\end{align}
Thus, the QW described in Eq.~\eqref{Z} is translation invariant. 

One can generalize the QW evolution \eqref{Z} by introducing time dependence of the coin operator $U$ along the evolution. The resulting propagation operator reads:
\begin{align}\label{Zt}
Z(n) = R \otimes U(n)\ket{\uparrow}\bra{\uparrow} + R^\dagger\otimes U(n)\ket{\downarrow}\bra{\downarrow},
\end{align}
where $n$ refers to the $n$-th step of the QW. This means the coin toss operator can change from one step to the next. Such time-dependent QWs  \cite{Banuls2006} preserve translation invariance \eqref{Z-translation-invariance}.

On the other hand, translation invariance can be broken if we generalize QW \eqref{Z} by introducing a shift $\Phi$ of the quasi-momentum,  given by the unitary 
\emph{ quasi-momentum shift} operation $E_\Phi$ on the walker:
\begin{align}\label{electric}
E_\Phi &= \sum_j e^{i\Phi j} \ket{j}\bra{j}.
\end{align}
For example, if the walker is a particle of unit charge, the shift $E_\Phi$ can be realized by a static  electric field $\Phi$, hence the name electric QW \cite{Wojcik2004,Romanelli2005,Genske2013,Cedzich2013}.
The propagation operator of an electric QW thus reads:
\begin{align}\label{ZE}
Z_{_E} &= (E_\Phi\otimes \mathbb{I})Z = E_\Phi R \otimes U\ket{\uparrow}\bra{\uparrow} + E_\Phi R^\dagger\otimes U\ket{\downarrow}\bra{\downarrow}.
\end{align}
It can be inspected that $Z_{_E}$ is not translation invariant, since it does not satisfy Eq.~\eqref{Z-translation-invariance}. 

There are a number of ways to define more complicated QWs. However, we focus here only on  three families: simple  QWs \eqref{Z}, QW with time dependent coin \eqref{Zt}, and  electric QWs \eqref{ZE}.
 Our goal is  to identify QWs with the same physical properties by considering unitary equivalence.

\section{Unitary equivalence}\label{unitary-equiv}

Two QW propagation operators, such as $Z$  in \eqref{Z} and 
\begin{align}\label{Zpr}
Z' &= (\mathbb{I}\otimes U')S',\nonumber\\
&= R \otimes U'\ket{\uparrow'}\bra{\uparrow'} + R^\dagger\otimes U'\ket{\downarrow'}\bra{\downarrow'}.
\end{align}
are unitary equivalent, if 
\begin{align}
Z' &= VZV^\dagger,
\end{align}
where $V$ is unitary. 
 A QW evolution under the action of $Z$ can be written as:
\begin{align}
\ket{\Psi(n)} &= Z^n \ket{\Psi(0)}= V^\dagger Z'^n V\ket{\Psi(0)}.
\end{align}
Thus, the dynamics of the initial state $\ket{\Psi(0)}$ under $Z$ is the same as the dynamics of the initial state $V\ket{\Psi(0)}$ under $Z'$. I.e., the statistics of the measurement of any observable $A$ after $n$ steps of a QW with initial state $\ket{\Psi(0)}$ and propagator $Z$ coincide with the statistics of a measurement of the rotated observable $V A V^\dagger$  in the system with initial state $V\ket{\Psi(0)}$ and $Z'$.

We can restrict the class of operators $V$ by demanding that certain properties of QWs are to be preserved. First of all, we request the entanglement between the coin and the walker to be not affected. Therefore, $V$ can only be of the form
\begin{align}\label{VXW}
V= W\otimes X
\end{align}
where the unitary operator $W$ acts on the walker states and $X\in SU(2)$ acts on the coin space. 
In addition, the product form of $V$ \eqref{VXW}  preserves the characteristic devision of simple QWs into a coin toss part and a conditional shift.  Optionally, we can require the translation invariance of $V$ which is fulfilled if $W$ is translation
invariant: 
\begin{align}\label{V-translation-invariance}
RWR^\dagger=W.
\end{align}
Or we can demand the invariance of the canonical basis $\{\ket{\uparrow},\ket{\downarrow}\}$,
meaning 
\begin{align}\label{canbasis-invariance}
X\ket{\uparrow}=\ket{\uparrow},~~~ X\ket{\downarrow}=\ket{\downarrow}.
\end{align}
Remarkably, a time dependent generalization of the unitary equivalence is worth of interest.
In particular, it is possible to establish an equivalence between the simple QW characterized by constant
$Z$ and the time-dependent QW characterized by time-dependent $Z(n)$:
\begin{align}
Z(n) &= V(n)ZV^\dagger(n),
\end{align}
where $V(n)= W(n)\otimes X(n)$ is time-dependent unitary. It leads to  equivalence between the electric QW and a  time-dependent QW. 

By means of  unitary equivalence, in the forthcoming subsection we introduce a trivial  canonical representation of simple QWs \eqref{Z},
we derive useful equivalence classes for them, and we point out the unitary equivalence of a particular time-dependent translation-invariant
QW with the electric QW.

\subsection{Canonical representation}
Since simple QWs \eqref{Z} are translation invariant, it is instructive 
to ensure translation invariance \eqref{V-translation-invariance} of the unitary 
operations $V$ of the equivalence transformations. We take the trivial choice $W=\mathbb{I}$, and consider transformations that act on the coin degree of freedom only. 
\begin{align}
V = \mathbb{I}\otimes X.
\end{align}
Consider the form of $U$ as given in Eq.\,\eqref{rotvec} which, on the Bloch-sphere,  corresponds to a $\phi$-rotation about the rotation axis $\vec{r}$. Let us choose a particular $X$ such that it rotates $\vec{r}$ to the vertical position:
\begin{align}
X &= \ket{\uparrow}\bra{\uparrow'} + \ket{\downarrow}\bra{\downarrow'},
\end{align}
where
\begin{align}
\vec{r}\cdot \vec{\sigma} \ket{\uparrow'} = \ket{\uparrow'}, \quad \vec{r}\cdot \vec{\sigma} \ket{\downarrow'} = -\ket{\downarrow'}.
\end{align}
This equivalence transformation yields the following single-parameter canonical form from the three-parameter coin toss operator \eqref{rotvec}:
\begin{align}
 U\to XUX^\dagger = \exp\left(\frac{i}{2}\phi\sigma_z\right),
\end{align}
at the price of yielding a two-parameter `tilted' canonical basis from the standard one:
 \begin{align}
\ket{\uparrow} \to  X\ket{\uparrow} = \ket{\uparrow'} ,\\
\ket{\downarrow} \to X\ket{\downarrow}  = \ket{\downarrow'}.
\end{align}
This transformation affects the conditional shift operator $S$ as follows:
\begin{align}
S \to & R \otimes \ket{\uparrow'}\bra{\uparrow'} + R^\dagger\otimes \ket{\downarrow'}\bra{\downarrow'}\,.
\end{align}
Thus,  the propagation operator $Z$ transforms under $V$ into its alternative canonical form:
\begin{align}
&Z \to  VZV^\dagger \\
             &= R \otimes\exp\left(\frac{i}{2}\phi\sigma_z\right)\ket{\uparrow'}\bra{\uparrow'} 
 + R^\dagger\otimes\exp\left(\frac{i}{2}\phi\sigma_z\right)\ket{\downarrow'}\bra{\downarrow'}\,.\nonumber
\end{align}
 Hence,  any simple QW of the form \eqref{Z}, can be realized by means of a relative phase shift (rotation about the $z$-axis)  as coin-toss operation, given that the subsequent shift of the walker is conditioned on a rotated basis. Moreover, 
this example shows how equivalence transformations of the coin-degree of freedom can be employed to find different realizations of the same quantum walk.

\subsection{Equivalence classes of simple QWs}\label{equiv-class}
In order to deduce the fundamental  structure of simple QWs \eqref{Z}, let us
find their partition into the largest non-trivial classes of unitary equivalent QWs.   For this purpose, we consider as above only those unitary transformations which preserve entanglement  and are thus of the form $V= W\otimes X$ \eqref{VXW}.  
However, contrary to the treatment in the last subsection we here restrict  to unitary operators  $X$ that 
do not alter the canonical basis, cf. Eq.~\eqref{canbasis-invariance}, but do not 
require the translation invariance \eqref{V-translation-invariance} for $W$.  

Consider the QW propagation operator $Z$ \eqref{Z} with the coin toss operator $U$ in Euler
parametrization \eqref{euler}:
\begin{align}\label{Z2}
Z_{\eta\theta\xi} =& R \otimes e^{i\frac{\eta}{2}\sigma_z}e^{i\frac{\theta}{2}\sigma_y}e^{i\frac{\xi}{2}}\ket{\uparrow}\bra{\uparrow} \nonumber\\
&+ R^\dagger\otimes e^{i\frac{\eta}{2}\sigma_z}e^{i\frac{\theta}{2}\sigma_y}e^{-i\frac{\xi}{2}}\ket{\downarrow}\bra{\downarrow},
\end{align}
where we have used the relations:
\begin{align}
e^{i\frac{\xi}{2}\sigma_z}\ket{\uparrow} = e^{i\frac{\xi}{2}}\ket{\uparrow},\qquad e^{i\frac{\xi}{2}\sigma_z}\ket{\downarrow} = e^{-i\frac{\xi}{2}}\ket{\downarrow}.
\end{align}
We choose $X$ as
\begin{align}\label{W}
X &= \exp\left(-i\frac{\eta}{2}\sigma_z\right),
\end{align}
and apply it to the coin part of the propagator $Z$:
\begin{align}\label{Z2}
&(I\otimes e^{-i\frac{\eta}{2}\sigma_z})Z_{\eta\theta\xi}(I\otimes  e^{i\frac{\eta}{2}\sigma_z})\\ 
&= R \otimes e^{i\frac{\theta}{2}\sigma_y}e^{i\frac{\xi+\eta}{2}}\ket{\uparrow}\bra{\uparrow}
 + R^\dagger\otimes e^{i\frac{\theta}{2}\sigma_y}e^{-i\frac{\xi+\eta}{2}}\ket{\downarrow}\bra{\downarrow}\nonumber,
\end{align}
The $(\xi+\eta)$-dependent  phase factors can be removed by a suitable quasi-momentum shift \eqref{electric}.
Observe that $E_\Phi R E_\Phi^\dagger =\exp(i\Phi)R$, hence we choose 
\begin{align}\label{electric-1b}
W &= E_{-(\eta+\xi)/2},
\end{align}
leading to
\begin{align}\label{ZtoZtheta}
(E_{-(\eta+\xi)/2}\otimes e^{-i\frac{\eta}{2}\sigma_z})Z_{\eta\theta\xi}( E_{(\xi+\eta)/2}\otimes  e^{i\frac{\eta}{2}\sigma_z})=Z_\theta 
\end{align}
where
\begin{align}\label{Ztheta}
Z_\theta =& R \otimes e^{i\frac{\theta}{2}\sigma_y}\ket{\uparrow}\bra{\uparrow}
 + R^\dagger\otimes e^{i\frac{\theta}{2}\sigma_y}\ket{\downarrow}\bra{\downarrow}.
\end{align}

Therefore, the three-parameter QW propagation operators $Z_{\theta\eta\xi}$ are unitary equivalent to the single-parameter $Z_\theta=Z_{0\theta 0}$. Only the middle Euler angle $\theta$ is relevant. The QW family $\{Z_\theta; \theta\in[0,\pi]\}$ 
universally represents all (simple) QWs.  Note, that the resulting coin toss  operator $e^{i\frac{\theta}{2}\sigma_y}$ corresponds to a real-valued matrix with respect to the canonical basis and  thus confirms the assumption that it means no restriction of generality to only consider coin-toss matrices with real entries \cite{Bach2004, Knight2004}.  
The angle $\theta$ in Eq.~\eqref{Z2} has already been shown to characterize the phase velocity and the group velocity in QW propagation \cite{Kempf2009}.

\subsection{Equivalence of time dependent and electric QWs}\label{time-electric-qw}
In this subsection we show that the electric QW \eqref{ZE} is equivalent to a particular time-dependent QW \eqref{Zt}. 
To show this equivalence first we note that the quasi-momentum shift $E_\Phi$ defined in \eqref{electric} satisfies the
identity $E_\Phi E_\chi=E_{\Phi+\chi}$. In particular, $E_\Phi$ can be rewritten as:
\begin{align}
E_\Phi &= E_{n\Phi}E_{(1-n)\Phi},\label{ET}
\end{align}
where $n$ is an integer to count the steps of the time dependent QW which we are going to construct. Using Eq.\,\eqref{ET} we can express the operator $E_\Phi R$ as:
\begin{align}
E_\Phi R &= e^{-i\Phi(n-1)} E_{n\Phi}R\,E_{(1-n)\Phi},
\end{align}
where we have used $E_{(1-n)\Phi}  R\, E_{(n-1)\Phi} = e^{-i\Phi(n-1)} R$. Thus  and without restriction of generality (cp.\ Subsec. \ref{equiv-class}), the propagation operator $Z_E =\left(E_\Phi \otimes \mathbb{I}\right) Z_\theta$ reads:
\begin{align}
Z_E =& E_{n\Phi}R\,E_{(1-n)\Phi} \otimes e^{i\frac{\theta}{2}\sigma_y}e^{-i\Phi(n-1)}\ket{\uparrow}\bra{\uparrow}\nonumber\\
& + E_{n\Phi}R^\dagger E_{(1-n)\Phi}\otimes e^{i\frac{\theta}{2}\sigma_y}e^{i\Phi(n-1)}\ket{\downarrow}\bra{\downarrow},
\end{align}
yielding 
\begin{align}
Z_E=& \left( E_{n\Phi}\otimes  \mathbb{I}\right) Z(n) \left(E_{(1-n)\Phi}\otimes  \mathbb{I}\right),
\label{ZE2}
\end{align}
where $Z(n)$ represents the time dependent quantum walk defined in \eqref{Zt} with 
\begin{align}
U(n) =& e^{i\frac{\theta}{2}\sigma_y}e^{-i(n-1)\Phi \sigma_z}
\end{align}
 as the time dependent coin operator.

In order to show the equivalence between the electric quantum walk and the time dependent quantum walk, let us consider the cumulative evolution operator after $1,\, 2,\, \cdots,\, n$ steps,
expressed in terms of the above time dependent QW:

\begin{align} 
 Z_E= & \left(E_{\Phi} \otimes \mathbb{I}\right)Z(1),\nonumber\\
Z_E^2 =& \left(E_{2\Phi} \otimes \mathbb{I}\right)Z(2)Z(1),\nonumber\\
 Z_E^3 =& \left(E_{3\Phi} \otimes \mathbb{I}\right)Z(3)Z(2)Z(1)\,\\
\dots\nonumber\\
 Z_E^n =& \left(E_{n\Phi} \otimes \mathbb{I}\right)Z(n)\dots Z(2)Z(1)\,\nonumber\\
\dots\nonumber
\end{align}
The expression for $n$ steps follows by induction from the one for $n-1$ steps by writing $Z_E^n= Z_E Z_E^{n-1}$ and inserting $Z_E$ as given in \eqref{ZE2}.  
Thus, the electric QW propagation is equivalent to the evolution with the time dependent coin 
where the stepwise electric kicks act cumulatively at a single time after the last step.

We note that the above equivalence is direct  and needs no unitary transformation on the one hand.
On the other hand, the equivalence is limited by the obligate cumulative electric kick.
Interestingly, we can get rid of this kick, at least  under special circumstances.   Suppose
the phase $\Phi$ is a rational multiple of $2\pi$, i.e., $\Phi = 2\pi p/q$, where $p$ and $q$ are integers, then
\begin{align}
\mathbb{I}= & E_{q\Phi} =E_{2q\Phi}=\dots= E_{nq\Phi} =\dots.
\end{align}
This results in the perfect equivalence between the $q$-step electric quantum walk and the $q$-step time dependent quantum walk:
\begin{align}
Z_E^{nq} = & \prod_{i=1}^{nq} Z(i),
\end{align}
for all $n$, with a time-ordering understood for the operator product.

\section{Conclusion and Discussion}\label{conc}
In this article we have applied the concept of unitary equivalence to one-dimensional QWs. For this purpose it was important to restrict to local unitaries that act independently on the coin and the walker degree of freedom in order to conserve the entanglement between both.  The restriction also limited our equivalence  to comprise only QW dynamics defined by the characteristic form of the QW propagator. The result can be used to find different realizations of the same dynamics on one hand, and to distinguish genuinely different dynamics on the other. Our central result is that the set of all three-parameter simple QWs is unitary equivalent with the family of single-parameter simple QWs. This will seriously simplify the discussion of
QW structure, e.g.,  the diagonalization of the evolution operator.
For example, it turns out that it is sufficient to consider the quantum walks with  real coin toss matrices to generate all possible distinct simple quantum walk dynamics. This result is important for the cases where one needs to scan the whole coin toss parameter space, for example, in studying the topological character of quantum walks \cite{Kitagawa2010}. Furthermore, the equivalence of particular quantum walks, such as electric QW and time dependent QW, provides an important insight into QW dynamics. 

The dynamics and the structure of two- or higher-dimensional quantum walk is more involved. Thus, equivalence classes might play an important role in the study of such evolutions. This will be the focus of future studies.
 
L.D. thanks the Hungarian Scientific Research Fund (Grant No. 75129) and the
EU COST Action MP1006 for support. This work is based on the research supported in part by  the  National  Research  Foundation  of  South  Africa  (Grant  specific  unique  reference  number  (UID)  86325).

%


\begin{thebibliography}{20}%
\makeatletter
\providecommand \@ifxundefined [1]{%
 \@ifx{#1\undefined}
}%
\providecommand \@ifnum [1]{%
 \ifnum #1\expandafter \@firstoftwo
 \else \expandafter \@secondoftwo
 \fi
}%
\providecommand \@ifx [1]{%
 \ifx #1\expandafter \@firstoftwo
 \else \expandafter \@secondoftwo
 \fi
}%
\providecommand \natexlab [1]{#1}%
\providecommand \enquote  [1]{``#1''}%
\providecommand \bibnamefont  [1]{#1}%
\providecommand \bibfnamefont [1]{#1}%
\providecommand \citenamefont [1]{#1}%
\providecommand \href@noop [0]{\@secondoftwo}%
\providecommand \href [0]{\begingroup \@sanitize@url \@href}%
\providecommand \@href[1]{\@@startlink{#1}\@@href}%
\providecommand \@@href[1]{\endgroup#1\@@endlink}%
\providecommand \@sanitize@url [0]{\catcode `\\12\catcode `\$12\catcode
  `\&12\catcode `\#12\catcode `\^12\catcode `\_12\catcode `\%12\relax}%
\providecommand \@@startlink[1]{}%
\providecommand \@@endlink[0]{}%
\providecommand \url  [0]{\begingroup\@sanitize@url \@url }%
\providecommand \@url [1]{\endgroup\@href {#1}{\urlprefix }}%
\providecommand \urlprefix  [0]{URL }%
\providecommand \Eprint [0]{\href }%
\providecommand \doibase [0]{http://dx.doi.org/}%
\providecommand \selectlanguage [0]{\@gobble}%
\providecommand \bibinfo  [0]{\@secondoftwo}%
\providecommand \bibfield  [0]{\@secondoftwo}%
\providecommand \translation [1]{[#1]}%
\providecommand \BibitemOpen [0]{}%
\providecommand \bibitemStop [0]{}%
\providecommand \bibitemNoStop [0]{.\EOS\space}%
\providecommand \EOS [0]{\spacefactor3000\relax}%
\providecommand \BibitemShut  [1]{\csname bibitem#1\endcsname}%
\let\auto@bib@innerbib\@empty
\bibitem [{\citenamefont {Aharonov}\ \emph {et~al.}(1993)\citenamefont
  {Aharonov}, \citenamefont {Davidovich},\ and\ \citenamefont
  {Zagury}}]{Aharonov1993}%
  \BibitemOpen
  \bibfield  {author} {\bibinfo {author} {\bibfnamefont {Y.}~\bibnamefont
  {Aharonov}}, \bibinfo {author} {\bibfnamefont {L.}~\bibnamefont
  {Davidovich}}, \ and\ \bibinfo {author} {\bibfnamefont {N.}~\bibnamefont
  {Zagury}},\ }\href@noop {} {\bibfield  {journal} {\bibinfo  {journal} {Phys.
  Rev. A}\ }\textbf {\bibinfo {volume} {48}},\ \bibinfo {pages} {1687}
  (\bibinfo {year} {1993})}\BibitemShut {NoStop}%
\bibitem [{\citenamefont {Nayak}\ and\ \citenamefont
  {Vishwanath}(2000)}]{Nayak2000}%
  \BibitemOpen
  \bibfield  {author} {\bibinfo {author} {\bibfnamefont {A.}~\bibnamefont
  {Nayak}}\ and\ \bibinfo {author} {\bibfnamefont {A.}~\bibnamefont
  {Vishwanath}},\ }\href@noop {} {\bibfield  {journal} {\bibinfo  {journal}
  {Arxiv preprint quant-ph/0010117}\ } (\bibinfo {year} {2000})}\BibitemShut
  {NoStop}%
\bibitem [{\citenamefont {Ambainis}\ \emph {et~al.}(2001)\citenamefont
  {Ambainis}, \citenamefont {Bach}, \citenamefont {Nayak}, \citenamefont
  {Vishwanath},\ and\ \citenamefont {Watrous}}]{Ambainis2001}%
  \BibitemOpen
  \bibfield  {author} {\bibinfo {author} {\bibfnamefont {A.}~\bibnamefont
  {Ambainis}}, \bibinfo {author} {\bibfnamefont {E.}~\bibnamefont {Bach}},
  \bibinfo {author} {\bibfnamefont {A.}~\bibnamefont {Nayak}}, \bibinfo
  {author} {\bibfnamefont {A.}~\bibnamefont {Vishwanath}}, \ and\ \bibinfo
  {author} {\bibfnamefont {J.}~\bibnamefont {Watrous}},\ }in\ \href@noop {}
  {\emph {\bibinfo {booktitle} {Proceedings of the thirty-third annual ACM
  symposium on Theory of computing}}}\ (\bibinfo {organization} {ACM},\
  \bibinfo {year} {2001})\ pp.\ \bibinfo {pages} {37--49}\BibitemShut {NoStop}%
\bibitem [{\citenamefont {Kempe}(2003)}]{Kempe2003}%
  \BibitemOpen
  \bibfield  {author} {\bibinfo {author} {\bibfnamefont {J.}~\bibnamefont
  {Kempe}},\ }\href@noop {} {\bibfield  {journal} {\bibinfo  {journal}
  {Contemporary Physics}\ }\textbf {\bibinfo {volume} {44:4}},\ \bibinfo
  {pages} {307} (\bibinfo {year} {2003})}\BibitemShut {NoStop}%
\bibitem [{\citenamefont {Chandrashekar}\ \emph {et~al.}(2008)\citenamefont
  {Chandrashekar}, \citenamefont {Srikanth},\ and\ \citenamefont
  {Laflamme}}]{Chandrashekar2008}%
  \BibitemOpen
  \bibfield  {author} {\bibinfo {author} {\bibfnamefont {C.~M.}\ \bibnamefont
  {Chandrashekar}}, \bibinfo {author} {\bibfnamefont {R.}~\bibnamefont
  {Srikanth}}, \ and\ \bibinfo {author} {\bibfnamefont {R.}~\bibnamefont
  {Laflamme}},\ }\href@noop {} {\bibfield  {journal} {\bibinfo  {journal}
  {Phys. Rev. A}\ }\textbf {\bibinfo {volume} {77}},\ \bibinfo {pages} {032326}
  (\bibinfo {year} {2008})}\BibitemShut {NoStop}%
\bibitem [{\citenamefont {Ambainis}(2003)}]{Ambainis2003}%
  \BibitemOpen
  \bibfield  {author} {\bibinfo {author} {\bibfnamefont {A.}~\bibnamefont
  {Ambainis}},\ }\href@noop {} {\bibfield  {journal} {\bibinfo  {journal}
  {International Journal of Quantum Information}\ }\textbf {\bibinfo {volume}
  {1}},\ \bibinfo {pages} {507} (\bibinfo {year} {2003})}\BibitemShut {NoStop}%
\bibitem [{\citenamefont {Shenvi}\ \emph {et~al.}(2003)\citenamefont {Shenvi},
  \citenamefont {Kempe},\ and\ \citenamefont {Whaley}}]{Shenvi2003}%
  \BibitemOpen
  \bibfield  {author} {\bibinfo {author} {\bibfnamefont {N.}~\bibnamefont
  {Shenvi}}, \bibinfo {author} {\bibfnamefont {J.}~\bibnamefont {Kempe}}, \
  and\ \bibinfo {author} {\bibfnamefont {K.}~\bibnamefont {Whaley}},\
  }\href@noop {} {\bibfield  {journal} {\bibinfo  {journal} {Phys. Rev. A}\
  }\textbf {\bibinfo {volume} {67}},\ \bibinfo {pages} {052307} (\bibinfo
  {year} {2003})}\BibitemShut {NoStop}%
\bibitem [{\citenamefont {Childs}\ and\ \citenamefont
  {Goldstone}(2004)}]{Childs2004}%
  \BibitemOpen
  \bibfield  {author} {\bibinfo {author} {\bibfnamefont {A.}~\bibnamefont
  {Childs}}\ and\ \bibinfo {author} {\bibfnamefont {J.}~\bibnamefont
  {Goldstone}},\ }\href@noop {} {\bibfield  {journal} {\bibinfo  {journal}
  {Phys. Rev. A}\ }\textbf {\bibinfo {volume} {70}},\ \bibinfo {pages} {022314}
  (\bibinfo {year} {2004})}\BibitemShut {NoStop}%
\bibitem [{\citenamefont {Lovett}\ \emph {et~al.}(2010)\citenamefont {Lovett},
  \citenamefont {Cooper}, \citenamefont {Everitt}, \citenamefont {Trevers},\
  and\ \citenamefont {Kendon}}]{Lovett2010}%
  \BibitemOpen
  \bibfield  {author} {\bibinfo {author} {\bibfnamefont {N.~B.}\ \bibnamefont
  {Lovett}}, \bibinfo {author} {\bibfnamefont {S.}~\bibnamefont {Cooper}},
  \bibinfo {author} {\bibfnamefont {M.}~\bibnamefont {Everitt}}, \bibinfo
  {author} {\bibfnamefont {M.}~\bibnamefont {Trevers}}, \ and\ \bibinfo
  {author} {\bibfnamefont {V.}~\bibnamefont {Kendon}},\ }\href@noop {}
  {\bibfield  {journal} {\bibinfo  {journal} {Phys. Rev. A}\ }\textbf {\bibinfo
  {volume} {81}},\ \bibinfo {pages} {042330} (\bibinfo {year}
  {2010})}\BibitemShut {NoStop}%
\bibitem [{\citenamefont {Mohseni}\ \emph {et~al.}(2008)\citenamefont
  {Mohseni}, \citenamefont {Rebentrost}, \citenamefont {Lloyd},\ and\
  \citenamefont {Aspuru-Guzik}}]{Mohseni2008}%
  \BibitemOpen
  \bibfield  {author} {\bibinfo {author} {\bibfnamefont {M.}~\bibnamefont
  {Mohseni}}, \bibinfo {author} {\bibfnamefont {P.}~\bibnamefont {Rebentrost}},
  \bibinfo {author} {\bibfnamefont {S.}~\bibnamefont {Lloyd}}, \ and\ \bibinfo
  {author} {\bibfnamefont {A.}~\bibnamefont {Aspuru-Guzik}},\ }\href@noop {}
  {\bibfield  {journal} {\bibinfo  {journal} {The Journal of chemical physics}\
  }\textbf {\bibinfo {volume} {129}},\ \bibinfo {pages} {174106} (\bibinfo
  {year} {2008})}\BibitemShut {NoStop}%
\bibitem [{\citenamefont {Bach}\ \emph {et~al.}(2004)\citenamefont {Bach},
  \citenamefont {Coppersmith}, \citenamefont {Goldschen}, \citenamefont
  {Joynt},\ and\ \citenamefont {Watrous}}]{Bach2004}%
  \BibitemOpen
  \bibfield  {author} {\bibinfo {author} {\bibfnamefont {E.}~\bibnamefont
  {Bach}}, \bibinfo {author} {\bibfnamefont {S.}~\bibnamefont {Coppersmith}},
  \bibinfo {author} {\bibfnamefont {M.~P.}\ \bibnamefont {Goldschen}}, \bibinfo
  {author} {\bibfnamefont {R.}~\bibnamefont {Joynt}}, \ and\ \bibinfo {author}
  {\bibfnamefont {J.}~\bibnamefont {Watrous}},\ }\href@noop {} {\bibfield
  {journal} {\bibinfo  {journal} {Journal of Computer and System Sciences}\
  }\textbf {\bibinfo {volume} {69}},\ \bibinfo {pages} {562 } (\bibinfo {year}{2004})}\BibitemShut {NoStop}%
\bibitem [{\citenamefont {Knight}\ and\ \citenamefont
  {Rold{\'a}n}(2004)}]{Knight2004}%
  \BibitemOpen
  \bibfield  {author} {\bibinfo {author} {\bibfnamefont {P.~L.}\ \bibnamefont
  {Knight}}\ and\ \bibinfo {author} {\bibfnamefont {E.}~\bibnamefont
  {Rold{\'a}n}},\ }\href@noop {} {\bibfield  {journal} {\bibinfo  {journal} {J.
  Mod. Opt.}\ }\textbf {\bibinfo {volume} {51}},\ \bibinfo {pages} {1761}
  (\bibinfo {year} {2004})}\BibitemShut {NoStop}%
\bibitem [{\citenamefont {Meyer}(1996)}]{Meyer1996B}%
  \BibitemOpen
  \bibfield  {author} {\bibinfo {author} {\bibfnamefont {D.~A.}\ \bibnamefont  {Meyer}}\ }\href@noop {} 
{\bibfield  {journal} {\bibinfo  {journal} {J. Stat. Phys.}\ }
\textbf {\bibinfo {volume} {85}},\ \bibinfo {pages} {551}  (\bibinfo {year} {1996})}\BibitemShut {NoStop}%
\bibitem [{\citenamefont {Meyer}(1996)}]{Meyer1996}%
  \BibitemOpen
  \bibfield  {author} {\bibinfo {author} {\bibfnamefont {D.~A.}\ \bibnamefont
  {Meyer}}\ }\href@noop {} {\bibfield  {journal} {\bibinfo  {journal} {Phys. Rev. E}\ }\textbf {\bibinfo {volume} {55}},\ \bibinfo {pages} {5261}
  (\bibinfo {year} {1996})}\BibitemShut {NoStop}%
\bibitem [{\citenamefont {W\'ojcik}\ \emph {et~al.}(2004)\citenamefont
  {W\'ojcik}, \citenamefont {\L{}uczak}, \citenamefont
  {Kurzy\ifmmode~\acute{n}\else \'{n}\fi{}ski}, \citenamefont {Grudka},\ and\
  \citenamefont {Bednarska}}]{Wojcik2004}%
  \BibitemOpen
  \bibfield  {author} {\bibinfo {author} {\bibfnamefont {A.}~\bibnamefont
  {W\'ojcik}}, \bibinfo {author} {\bibfnamefont {T.}~\bibnamefont {\L{}uczak}},
  \bibinfo {author} {\bibfnamefont {P.}~\bibnamefont
  {Kurzy\ifmmode~\acute{n}\else \'{n}\fi{}ski}}, \bibinfo {author}
  {\bibfnamefont {A.}~\bibnamefont {Grudka}}, \ and\ \bibinfo {author}
  {\bibfnamefont {M.}~\bibnamefont {Bednarska}},\ }\href@noop {} {\bibfield
  {journal} {\bibinfo  {journal} {Phys. Rev. Lett.}\ }\textbf {\bibinfo
  {volume} {93}},\ \bibinfo {pages} {180601} (\bibinfo {year}
  {2004})}\BibitemShut {NoStop}%
\bibitem [{\citenamefont {Romanelli}\ \emph {et~al.}(2005)\citenamefont
  {Romanelli}, \citenamefont {Auyuanet}, \citenamefont {Siri}, \citenamefont
  {Abal},\ and\ \citenamefont {Donangelo}}]{Romanelli2005}%
  \BibitemOpen
  \bibfield  {author} {\bibinfo {author} {\bibfnamefont {A.}~\bibnamefont
  {Romanelli}}, \bibinfo {author} {\bibfnamefont {A.}~\bibnamefont {Auyuanet}},
  \bibinfo {author} {\bibfnamefont {R.}~\bibnamefont {Siri}}, \bibinfo {author}
  {\bibfnamefont {G.}~\bibnamefont {Abal}}, \ and\ \bibinfo {author}
  {\bibfnamefont {R.}~\bibnamefont {Donangelo}},\ }\href@noop {} {\bibfield
  {journal} {\bibinfo  {journal} {Physica A: Statistical Mechanics and its
  Applications}\ }\textbf {\bibinfo {volume} {352}},\ \bibinfo {pages} {409 }
  (\bibinfo {year} {2005})}\BibitemShut {NoStop}%
\bibitem [{\citenamefont {Genske}\ \emph {et~al.}(2013)\citenamefont {Genske},
  \citenamefont {Alt}, \citenamefont {Steffen}, \citenamefont {Werner},
  \citenamefont {Werner}, \citenamefont {Meschede},\ and\ \citenamefont
  {Alberti}}]{Genske2013}%
  \BibitemOpen
  \bibfield  {author} {\bibinfo {author} {\bibfnamefont {M.}~\bibnamefont
  {Genske}}, \bibinfo {author} {\bibfnamefont {W.}~\bibnamefont {Alt}},
  \bibinfo {author} {\bibfnamefont {A.}~\bibnamefont {Steffen}}, \bibinfo
  {author} {\bibfnamefont {A.~H.}\ \bibnamefont {Werner}}, \bibinfo {author}
  {\bibfnamefont {R.~F.}\ \bibnamefont {Werner}}, \bibinfo {author}
  {\bibfnamefont {D.}~\bibnamefont {Meschede}}, \ and\ \bibinfo {author}
  {\bibfnamefont {A.}~\bibnamefont {Alberti}},\ }\href@noop {} {\bibfield
  {journal} {\bibinfo  {journal} {Phys. Rev. Lett.}\ }\textbf {\bibinfo
  {volume} {110}},\ \bibinfo {pages} {190601} (\bibinfo {year}
  {2013})}\BibitemShut {NoStop}%
\bibitem [{\citenamefont {Cedzich}\ \emph {et~al.}(2013)\citenamefont
  {Cedzich}, \citenamefont {Ryb\'ar}, \citenamefont {Werner}, \citenamefont
  {Alberti}, \citenamefont {Genske},\ and\ \citenamefont
  {Werner}}]{Cedzich2013}%
  \BibitemOpen
  \bibfield  {author} {\bibinfo {author} {\bibfnamefont {C.}~\bibnamefont
  {Cedzich}}, \bibinfo {author} {\bibfnamefont {T.}~\bibnamefont {Ryb\'ar}},
  \bibinfo {author} {\bibfnamefont {A.~H.}\ \bibnamefont {Werner}}, \bibinfo
  {author} {\bibfnamefont {A.}~\bibnamefont {Alberti}}, \bibinfo {author}
  {\bibfnamefont {M.}~\bibnamefont {Genske}}, \ and\ \bibinfo {author}
  {\bibfnamefont {R.~F.}\ \bibnamefont {Werner}},\ }\href@noop {} {\bibfield
  {journal} {\bibinfo  {journal} {Phys. Rev. Lett.}\ }\textbf {\bibinfo
  {volume} {111}},\ \bibinfo {pages} {160601} (\bibinfo {year}
  {2013})}\BibitemShut {NoStop}%
\bibitem [{\citenamefont {Ba\~nuls}\ \emph {et~al.}(2006)\citenamefont
  {Ba\~nuls}, \citenamefont {Navarrete}, \citenamefont {P\'erez}, \citenamefont
  {Rold\'an},\ and\ \citenamefont {Soriano}}]{Banuls2006}%
  \BibitemOpen
  \bibfield  {author} {\bibinfo {author} {\bibfnamefont {M.~C.}\ \bibnamefont
  {Ba\~nuls}}, \bibinfo {author} {\bibfnamefont {C.}~\bibnamefont {Navarrete}},
  \bibinfo {author} {\bibfnamefont {A.}~\bibnamefont {P\'erez}}, \bibinfo
  {author} {\bibfnamefont {E.}~\bibnamefont {Rold\'an}}, \ and\ \bibinfo
  {author} {\bibfnamefont {J.~C.}\ \bibnamefont {Soriano}},\ }\href@noop {}
  {\bibfield  {journal} {\bibinfo  {journal} {Phys. Rev. A}\ }\textbf {\bibinfo
  {volume} {73}},\ \bibinfo {pages} {062304} (\bibinfo {year}
  {2006})}\BibitemShut {NoStop}%
\bibitem [{Note1()}]{Note1}%
  \BibitemOpen
  \bibinfo {note} {By simple we mean a translationally invariant and
  time-independent quantum.}\BibitemShut {Stop}%
\bibitem [{\citenamefont {Kempf}\ and\ \citenamefont
  {Portugal}(2009)}]{Kempf2009}%
  \BibitemOpen
  \bibfield  {author} {\bibinfo {author} {\bibfnamefont {A.}~\bibnamefont
  {Kempf}}\ and\ \bibinfo {author} {\bibfnamefont {R.}~\bibnamefont
  {Portugal}},\ }\href@noop {} {\bibfield  {journal} {\bibinfo  {journal}
  {Phys. Rev. A}\ }\textbf {\bibinfo {volume} {79}},\ \bibinfo {pages} {052317}
  (\bibinfo {year} {2009})}\BibitemShut {NoStop}%
\bibitem [{\citenamefont {Kitagawa}\ \emph {et~al.}(2010)\citenamefont
  {Kitagawa}, \citenamefont {Rudner}, \citenamefont {Berg},\ and\ \citenamefont
  {Demler}}]{Kitagawa2010}%
  \BibitemOpen
  \bibfield  {author} {\bibinfo {author} {\bibfnamefont {T.}~\bibnamefont
  {Kitagawa}}, \bibinfo {author} {\bibfnamefont {M.~S.}\ \bibnamefont
  {Rudner}}, \bibinfo {author} {\bibfnamefont {E.}~\bibnamefont {Berg}}, \ and\
  \bibinfo {author} {\bibfnamefont {E.}~\bibnamefont {Demler}},\ }\href@noop {}
  {\bibfield  {journal} {\bibinfo  {journal} {Phys. Rev. A}\ }\textbf {\bibinfo
  {volume} {82}},\ \bibinfo {pages} {033429} (\bibinfo {year}
  {2010})}\BibitemShut {NoStop}%
\end{thebibliography}
\end{document}